\documentclass[fleqn,usenatbib]{mnras}

\usepackage{newtxtext,newtxmath}
\usepackage[T1]{fontenc}
\usepackage{ae,aecompl}

\usepackage{graphicx}	% Including figure files
\usepackage{amsmath}	% Advanced maths commands
\usepackage{slashed}    % Slashed operators
\DeclareMathOperator*{\argminT}{argmin}
\DeclareMathOperator*{\argmaxT}{argmax}
\DeclareMathOperator*{\minT}{min}
\DeclareMathOperator*{\maxT}{max}
\usepackage{mathtools}
\DeclarePairedDelimiterX{\norm}[1]{\lVert}{\rVert}{#1}
\usepackage{wrapfig}
\newcommand{\emdash}{\nobreak--\nobreak\hskip4pt}
\usepackage{bm}
\makeatletter
\newcommand*{\rom}[1]{\expandafter\@slowromancap\romannumeral #1@}
\makeatother
\usepackage{multirow}
\usepackage{bbold}

\usepackage{smartdiagram}
\usepackage{tikz}
\usetikzlibrary{shapes,arrows}
\tikzstyle{block} = [rectangle, draw, fill=blue!10,
    text width=12em, text centered, rounded corners, minimum height=3em]
\tikzstyle{line} = [draw, -latex']
\tikzstyle{decision} = [diamond, draw, fill=purple!10,
    text width=6.0em, text badly centered, node distance=2.2cm, inner sep=0pt]

\newcommand{\vect}[1]{\boldsymbol{#1}}

\usepackage{algorithm}
\usepackage{setspace}
\usepackage[noend]{algpseudocode}
\makeatletter
\def\BState{\State\hskip-\ALG@thistlm}
\makeatother

\title[Mass-mapping with uncertainty quantification]{Sparse Bayesian mass-mapping with uncertainties: local credible intervals}

\author[ Price et al.]{
M. A.Price$^{1}$\thanks{E-mail: m.price.17@ucl.ac.uk}, X. Cai$^{1}$, J. D. McEwen$^{1}$, M. Pereyra$^{2}$, T. D. Kitching$^{1}$ \newauthor
\normalsize(for the LSST Dark Energy Science Collaboration)
\\
$^{1}$Mullard Space Science Laboratory, University College London, RH5 6NT, UK.\\
$^{2}$Maxwell Institute for Mathematical Sciences, Heriot-Watt University, Edinburgh EH14 4AS, United Kingdom \\
}

\date{Accepted XXX. Received YYY; in original form ZZZ}

\pubyear{2018}

\begin{document}
\label{firstpage}
\pagerange{\pageref{firstpage}--\pageref{lastpage}}
\maketitle

\begin{abstract}
Until recently mass-mapping techniques for weak gravitational lensing convergence reconstruction have lacked a principled statistical framework upon which to quantify reconstruction uncertainties, without making strong assumptions of Gaussianity. In previous work we presented a sparse hierarchical Bayesian formalism for convergence reconstruction that addresses this shortcoming. Here, we draw on the concept of \textit{local credible intervals} (\textit{cf.} Bayesian error bars) as an extension of the uncertainty quantification techniques previously detailed. These uncertainty quantification techniques are benchmarked against those recovered \textit{via} Px-MALA \emdash a state of the art proximal Markov Chain Monte Carlo (MCMC) algorithm. We find that typically our recovered uncertainties are everywhere conservative (never underestimate the uncertainty, yet the approximation error is bounded above), of similar magnitude and highly correlated with those recovered \textit{via} Px-MALA. Moreover, we demonstrate an increase in computational efficiency of $\mathcal{O}(10^6)$ when using our sparse Bayesian approach over MCMC techniques. This computational saving is critical for the application of Bayesian uncertainty quantification to large-scale stage \rom{4} surveys such as LSST and Euclid.
\end{abstract}


%%%%%%%%%%%%%%%%%%%%%%%%%%%%%%%%%%%%%%%%%%%%%%%%%%

%%%%%%%%%%%%%%%%% BODY OF PAPER %%%%%%%%%%%%%%%%%%
\section{Introduction} \label{sec:introduction}

As photons from distant sources (galaxies) travel through spacetime to us here and now their trajectories are perturbed by local mass over and under-densities, causing the observed shapes of structures to be warped, or \textit{gravitationally lensed}. This cosmological effect is sensitive to all matter (both visible and invisible), and so provides a natural cosmological probe of dark matter.
\par
The gravitational lensing effect has (at first order) two distinct effects: distant galaxies are magnified by a convergence field $\kappa$; and the third-flattening (ellipticity) is perturbed from an underlying intrinsic value by a shear field $\gamma$. A wide range of cosmology can be extracted from just the shear field \citep{[49],[42]}, though increasingly higher order statistics \citep{[23],[24],[27],[28]} are being computed on convergence maps directly.
\par
As a result of the mass-sheet degeneracy \citep[an \textit{a priori} degeneracy of the intrinsic brightness of galaxies, see][]{[1]} the convergence field cannot be observed directly. Instead measurements of the shear field $\gamma$ must be taken and inverted through some mapping to create an estimator for $\kappa$. Typically, these inverse problems are ill-posed (often seriously) and so creating unbiased estimators for the convergence $\kappa$ can prove difficult.
\par
Many convergence inversion techniques have been considered \citep[\textit{e.g.}][]{[29],[6],[3],[15],[22]} though the simplest, most direct method in the planar setting is that of Kaiser-Squires (KS) inversion \citep{[5]}. Though these methods often produce reliable estimates of $\kappa$, they all either lack principled statistical uncertainties on their reconstructions or make strong assumptions of Gaussianity (which heavily degrades the quality of non-Gaussian information in particular).
\par
For example, Wiener filtering \citep[see \textit{e.g}][]{Horowitz2018} directly adopts Gaussian priors which are more explicit assumptions of Gaussianity.  Other approaches, such as the KS method recover a noisy convergence estimate which is post-processed \textit{via} convolution with a Gaussian kernel, which promotes Gaussianity.
\par
On large scales the lensing information is primarly Gaussian in nature, though on smaller scales (at higher resolutions) there becomes a non-neglibile non-Gaussian contribution which encodes information about baryonic interactions and clustering amongst other non-linear effects. Analysis of such effects is expected \citep{[23]} to provide competitive and more importantly complementary constraints on cosmological parameters -- in particular parameters related closely to
  dark matter such as $\sigma_8$ and $\Omega_M$. Consequently, mapping techniques which preserve the non-Gaussian information content are a crucial step forward for dark matter analysis \textit{via} weak gravitational lensing.
\par
In previous work \citep{[M1]} we presented a new sparse hierarchical Bayesian formalism for reconstructing the convergence field. This not only regularizes the ill-posed inverse problem but allows us to explore the Bayesian posterior in order to recover principled uncertainties on our reconstruction.
It is important to note here that this mathematical framework is entirely general and can be applied for any posterior which belongs to the set of log-concave functions -- of which both sparsity enforcing Laplace type priors and standard Gaussian priors are members.
\par
Often hierarchical Bayesian inference problems are solved by \textit{Markov Chain Monte Carlo} (MCMC) techniques \citep[see \textit{e.g.}][]{[46]}, which explicitly return a large number of samples from the full posterior distribution \emdash from which one can construct true Bayesian uncertainties. Samples of the posterior \textit{via} MCMC algorithms construct theoretically optimal estimates of the posterior (in the limit of a large number of samples), but in practice can be extremely computationally taxing to recover fully.
\par
In fact, when the dimensionality becomes large these methods become infeasible \emdash often referred to as \textit{the curse of dimensionality}. In the context of lensing inverse problems each pixel constitutes a dimension, and so for a resolution of $1024\times1024$ (which is typical) the dimension of the problem is $\mathcal{O}(10^6)$.
\par
Recent advancements in probability density theory \citep{[19]} allow conservative approximations of Bayesian credible regions of the posterior from knowledge of the MAP solution alone \citep{[10]}. The sparse Bayesian method presented in previous work \citep[see][]{[M1]} recasts the maximization of the posterior distribution as a convex optimization problem from which the \textit{maximum a posteriori} (MAP) solution can be rapidly computed. Uncertainty quantification is then conducted utilizing the aforementioned approximate credible regions of the posterior.
In \citet{[M1]} hypothesis testing (determining the statistical significance of a feature of the recovered convergene map) was introduced to the weak lensing setting as a form of uncertainty quantification.
\par
In this article we introduce a further uncertainty quantification technique called \textit{local credible intervals} (\textit{cf.} pixel-level error bars).  Both hypothesis testing and local credible intervals were previously developed and applied to the radio interferometric setting \citep{[11],[12]}.  We also remark that there are alternative ways of testing image structures \citep{RPW18}. This paper serves as a benchmark comparison of our sparse hierarchical Bayesian formalism \citep[see][]{[M1]} to a bespoke MCMC algorithm, Px-MALA \citep{[11],[12],[40],[41]}. Px-MALA utilizes Moreau-Yoshida envelopes and proximity operators (tools from convex analyses) to support non-differentiable terms in the prior or likelihood, making it one of the only somewhat efficient ways to support  non-smooth sparsity-promoting priors (on which our sparse Bayesian mass-mapping framework is based) in high dimensional settings.
\par
The remainder of this article is structured as follows. We begin with section \ref{sec:HierarchicalBayesianInference} in which we review our sparse hierarchical Bayesian models for mass-mapping and present a brief overview of the Px-MALA MCMC algorithm. We then cover the relevant mathematical background of approximate Bayesian uncertainty quantification in section \ref{sec:MAPUncertainties} before introducing the concept of \textit{local credible intervals} \emdash an additional form of uncertainty quantification. In section \ref{sec:Testing}, we conduct a series of mock scenarios to compare the uncertainties recovered by our \textit{maximum a posteriori} (MAP) approach, and the full MCMC (Px-MALA) treatment. Finally we draw conclusions and discuss future work in section \ref{sec:Conclusions}.
\par
Section \ref{sec:HierarchicalBayesianInference} relies on a strong understanding of Bayesian inference and MCMC techniques along with a moderate understanding of proximal calculus and compressed sensing. As such, for the reader interested only in the application and benchmarking section \ref{sec:Testing} onwards is relevant content.

\section{Hierarchical Bayesian Inference for Mass-mapping} \label{sec:HierarchicalBayesianInference}
Hierarchical Bayesian models provide a flexible, well defined approach for dealing with uncertainties in a variety of problems. For an overview of Bayesian hierarchical modeling and MCMC techniques in the context of astrophysics we refer the reader to \citet{[46]}.
\par
We begin by presenting an overview of the sparse hierarchical Bayesian approach developed in previous work \citep[see][]{[M1]}, where we also review the weak lensing planar forward model. Following this we make the MAP optimization problem explicit.
We then review the Bayesian parameter inference hierarchy adopted in our sparse Bayesian mass-mapping algorithm \citep{[M1]}. Finally we provide a short introduction to the Px-MALA and MYULA proximal Markov chain Monte-Carlo algorithms \citep{[40],[41]}.

\subsection{Bayesian Inference} \label{sec:BayesianInference}
Mathematically, let us begin by considering the \textit{posterior distribution} which by Bayes' Theorem is given by
\begin{equation} \label{eq:bayes}
p(\kappa|\gamma) = \frac{p(\gamma|\kappa)p(\kappa)}{\int_{\mathbb{C}^N} p(\gamma|\kappa)p(\kappa)d\kappa}.
\end{equation}
Bayes' theorem relates the posterior distribution $p(\kappa|\gamma)$ to the product of some likelihood function $p(\gamma|\kappa)$ and some prior $p(\kappa)$. It is important to note here that a model is implicit which collectively defines the noise and the proposed relationship between observations $\gamma$ and inferences $\kappa$ \emdash specifically this term characterizes the noise model and the assumed mapping $\lbrace \kappa \mapsto \gamma \rbrace$. Note that the denominator in equation \eqref{eq:bayes} is the model's marginal likelihood which is unrelated to $\kappa$.
\par
Suppose the discretized complex shear field $\gamma \in \mathbb{C}^M$ and the discretized complex convergence field $\kappa \in \mathbb{C}^N$ \emdash where $M$ represents the number of binned shear measurements and $N$ represents the dimensionality of the convergence estimator \emdash are related by a measurement operator $\bm{\Phi} \in \mathbb{C}^{M \times N}$ defined such that
\begin{equation}
\bm{\Phi} \in \mathbb{C}^{M \times N} : \kappa \in \mathbb{C}^N \mapsto \gamma \in \mathbb{C}^M.
\end{equation}
Further, suppose a contaminating noise $n$ is present. Measurements of $\gamma$ are produced \textit{via}
\begin{equation} \label{eq:aquired_measurements}
\gamma = \bm{\Phi} \kappa + n.
\end{equation}
For the case considered within this paper, we take $n \sim\mathcal{N}(0,\sigma_n^2) \in \mathbb{C}^M$ \emdash \textit{i.e.} i.i.d. (independent and identically distributed) additive Gaussian noise. For the purpose of this paper we consider the simplest planar mapping,
\begin{equation} \label{eq:measurement_operator}
\bm{\Phi} = \bm{\mathsf{F}}^{-1} \bm{\mathsf{D}} \bm{\mathsf{F}}.
\end{equation}
Here, $\bm{\mathsf{F}}$ ($\bm{\mathsf{F}}^{-1}$) is the forward (inverse) discrete fast Fourier transforms and $\bm{\mathsf{D}}$ is the weak lensing planar forward-model in Fourier space \citep[\textit{e.g.}][]{[5]},
\begin{equation} \label{eq:ks}
\bm{\mathsf{D}}_{k_x,k_y} = \frac{k_x^2-k_y^2+2ik_xk_y}{k_x^2+k_y^2}.
\end{equation}
The measurement operator $\bm{\Phi}$ has also been extended to super-resolution image recovery \citep{[M1]}, but that is beyond the scope of this paper.
\par
In the majority of weak lensing surveys $M < N$ (\textit{i.e.} the shear field is a discrete under-sampling of the underlying convergence field) and so inverting the forward-model is typically ill-posed (often seriously). To regularize ill-posed inverse problems a term encoding prior information is introduced -- this is referred to either as the prior or regularization term.
\par
We choose a prior which reflects the quasi-philosophical  notion of \textit{Occam's Razor} \emdash a prior which says if two solutions are equally viable, the one which makes the fewest assumptions (the fewest active variables \emdash non-zero coefficients in a sparse domain) is more likely to be true. Mathematically, this is equivalent to imposing sparsity that minimizes the number of non-zero coefficients in a sparse representation (dictionary).
\par
One could select any sparsifying domain, though a natural choice for most physical systems are wavelets. We choose to use wavelets as our sparsifying dictionary in this paper and in previous work.
\par
The natural sparsity-promoting prior is the $\ell_0$-norm $\norm{.}_0$, often referred to as the \textit{Hamming distance} \emdash \textit{i.e.} the total number of non-zero coefficients of a field. However, this function is non-differentiable and (perhaps more importantly) non-convex. As such it cannot exploit the computational advantages provided by conventional convex optimization techniques.
\par
Researchers therefore often select the next most natural sparsity-promoting prior, the $\ell_1$-norm $\norm{.}_1$, which is convex and can be shown to share the same MAP (maximum-a-posteriori) solution as if one were to use the $\ell_0$-norm in certain cases \citep[see \textit{e.g.}][on \textit{convex relaxation}]{Donoho2006, Candes2008}.
\par
We now define the likelihood function (data fidelity term) as a multivariate Gaussian with diagonal covariance $\Sigma = \sigma_n^2 \mathbb{I}$ such that,
\begin{equation}
p(\gamma|\kappa) \propto \exp \Bigg(\frac{-\norm{\bm{\Phi} \kappa - \gamma}_2^2}{2\sigma_n^2} \Bigg),
\end{equation}
which \citep[as in][]{[M1]} is regularized by a non-differentiable Laplace-type sparsity-promoting wavelet prior
\begin{equation} \label{eq:l1-prior}
p(\kappa) \propto \exp \Big(-\mu \norm{\bm{\Psi}^{\dag}\kappa}_1 \Big),
\end{equation}
where $\bm{\Psi}$ is an appropriately selected sparsifying dictionary (such as a wavelet dictionary) in which the signal is assumed to be sparse, and $\mu \in \mathbb{R}_{+}$ is a regularization parameter. Substituting $p(\gamma|\kappa)$ and $p(\kappa)$ into equation (\ref{eq:bayes}) yields
\begin{equation} \label{eq:bayes-mm}
p(\kappa|\gamma) \propto \exp  \Bigg\{ - \Bigg( \mu \norm{ \bm{\Psi}^{\dag}\kappa}_1 + \frac{\norm{\bm{\Phi} \kappa - \gamma}_2^2}{2\sigma_n^2} \Bigg)  \Bigg\}.
\end{equation}
Note that one can choose any convex log-priors \textit{e.g.} an $\ell_2$-norm prior from which one essentially recovers Weiner filtering \citep[see][for alternate iterative Weiner filtering approaches]{Seljak2003,Horowitz2018}.

\subsection{Sparse MAP estimator} \label{sec:SparseMAP}
Drawing conclusions directly from $p(\kappa|\gamma)$ can be difficult because of the high dimensionality involved, which will be detailed in the next section. As an alternative, Bayesian methods often derive solutions by computing estimators that summarize $p(\kappa|\gamma)$, such as
maximizing the probability of the recovered $\kappa$ conditional on the data $\gamma$. Such a solution is referred to as the MAP solution. From the monotonicity of the logarithm function it is evident that,
\begin{align} \label{eq:log-posterior}
\begin{split}
\kappa^{\text{map}}  & = \argmaxT_{\kappa} \big \lbrace p(\kappa|\gamma) \big \rbrace  \\
				    & =  \argminT_{\kappa} \big \lbrace -\log ( \; p(\kappa|\gamma) \;) \big \rbrace \\
				    & = \argminT_{\kappa} \Bigg \lbrace \underbrace{ \mu \norm{ \bm{\Psi}^{\dag}\kappa}_1 }_{f(\kappa)} + \underbrace{ {\norm{\bm{\Phi} \kappa - \gamma}_2^2}/{2\sigma_n^2} }_{g(\kappa)} \Bigg \rbrace,
\end{split}
\end{align}
which is a convex minimization problem and can therefore be computed in a highly computationally efficient manner.
\par
To solve the convex minimization problem given in equation (\ref{eq:log-posterior}) we implement an adapted forward-backward splitting algorithm \citep{[45]}. A complete description of the steps adopted when solving this optimization problem, and the full details of the sparse hierarchical Bayesian formalism are outlined in previous work \citep{[12],[M1]}.

\subsection{Sparse Dictionary and Regularization Parameter} \label{sec:RegularizationSelection}
Here we provide a concise overview of the parameter selection aspect of our sparse Bayesian mass-mapping algorithm which was developed and presented in previous work \emdash for a complete description see \citet{[M1],[16]}.
\par
The prior term in equation (\ref{eq:log-posterior}) promotes the \textit{a priori} knowledge that the signal of interest $\kappa$ is likely to be sparse in a given dictionary $\bm{\Psi}$. A function $f(x)$ is sparse in a given dictionary $\bm{\Psi}$ if the number of non-zero coefficients is small compared to the total size of the dictionary domain. Wavelets form a general set of naturally sparsifying dictionaries for a wide-range of physical problems \emdash and have recently been shown to work well in the weak lensing setting \citep{[15],[6],[9],[M1]}.
For the purpose of this paper we restrict ourselves to Daubechies 8 (DB8) wavelets (with 8 wavelet levels) for simplicity though a wide variety of wavelets could be considered \citep[\textit{e.g.}][]{[39],[17],[18]}.
Note that the exact choice of wavelet representation (and prior) is independent from the results of this benchmarking paper, thus discussion of dictionary (and prior) optimality is not
of primary concern.
\par
An issue in these types of regularized optimization problems is the setting of regularization parameter $\mu$ - several approaches have been presented \citep{[6], [9], [14], [15]}. For uncertainties on reconstructed $\kappa$ maps to be truly principled $\mu$ must be computed in a well defined, statistically principled way. In \citet{[M1]} a hierarchical Bayesian inference approach to compute the theoretically optimal $\mu$ was adopted, which we outline in appendix \ref{sec:MM}

\subsection{Proximal MCMC Sampling} \label{sec:MCMC}
Sampling a full posterior distribution is very challenging in high dimensional settings, particularly when the prior $p(\kappa)$ considered is non-differentiable \emdash like the sparsity-promoting prior given in equation (\ref{eq:l1-prior}). In the following, we recall two proximal MCMC methods developed in \citet{[40],[41]} \emdash MYULA and Px-MALA \emdash  which can be applied to sample the full posterior density $p(\kappa|\gamma)$ for mass-mapping. After a set of samples has been obtained, various kinds of analysis can be performed, such as summary estimators of $\kappa$, and a range of uncertainty quantification techniques, as presented in \cite{[11],[12]}.
\par
For a probability density $p \in \mathcal{C}^1$ with Lipschitz gradient, the Markov chain of the unadjusted Langevin algorithm (ULA) to generate a set of samples $\{{\vect l}^{(m)}\}_{\in \mathbb{C}^N}$ based on a forward
Euler-Maruyama approximation with step-size $\delta > 0$ has the form
\begin{equation} \label{eqn:ldp-d}
{\vect l}^{(m+1)} = {\vect l}^{(m)} + \frac{\delta}{2} \nabla \log  {p}[{\vect l}^{(m)}] + \sqrt{\delta} {\vect w}^{(m+1)},
\end{equation}
where ${\vect w}^{(m+1)} \sim {\cal N} (0,\mathbb{1}_N)$ (an $N$-sequence of standard Gaussian random variables).
\par
However, the chain generated by ULA given above converges to ${p}$ with asymptotic bias. This kind of bias can be corrected at the expense of some additional estimation variance \citep{RT96} after involving a Metropolis-Hasting (MH) accept-reject step in ULA, which results in the MALA algorithm (Metropolis-adjusted Langevin Algorithm). However, the convergence of  ULA and MALA is limited to a continuously differentiable $\log{p}$ with Lipschitz gradient, which prohibits their application to our focus on mass-mapping with non-differentiable sparsity-promoting prior in equation (
\ref{eq:l1-prior}).
\par
Proximal MCMC methods \emdash such as MYULA and Px-MALA \emdash can be used to address non-differentiable sparsity-promoting priors \citep{[40],[41]}. Without loss of generality, consider a log-concave posterior which is of the exponential family
\begin{equation}
p(\kappa | \gamma) \propto \exp{\{-f(\kappa) -g(\kappa)\}},
\end{equation}
for lower semi-continuous convex and Lipschitz differentiable log-likelihood $g(x) \in \mathcal{C}^1$ and lower semi-continuous convex log-prior $f(x) \notin \mathcal{C}^1$. It is worth noting that this is precisely the setting adopted within this paper, where from \eqref{eq:bayes-mm}
\begin{equation}
f(\kappa) =  \mu \norm{ \bm{\Psi}^{\dag}\kappa}_1, \quad \text{and} \quad g(\kappa) = {\norm{\bm{\Phi} \kappa - \gamma}_2^2}/{2\sigma_n^2}.
\end{equation}
\par
To sample this posterior the gradient $\nabla \log p$ is required, however $f(x)$ is not Lipschitz differentiable. To account for the non-differentiability of $f(x)$ let us now define the smooth approximation $p_\lambda(\kappa|\gamma) \propto \exp{\{-f^\lambda(\kappa) -g(\kappa)\}}$, where
\begin{equation}
f^{\lambda} ({\kappa})  \equiv \min_{\hat{\kappa}\in \mathbb{C}^N}  \left \{ f(\hat{\kappa}) + \|\hat{\kappa} - {\kappa}\|^2/2\lambda \right \},
\end{equation}
is the $\lambda$-Moreau-Yosida envelope of $f$, which can be made arbitrarily close to $f$ by letting $\lambda \rightarrow 0$ (see \citealt{PB14}).
Then we have $ \underset{\lambda\rightarrow 0}{\lim} p_\lambda(\kappa | \gamma) = p(\kappa | \gamma)$, and more importantly that, for any $\lambda > 0$, the total-variation distance between the distributions $p_\lambda $ and $p$ is bounded by $\|p_\lambda - p\|_{TV} \leq \lambda \mu N$, providing an explicit bound on the estimation errors involved in using $p_\lambda$ instead of $p$ (see \citealt{[40]} for details). Also, the gradient $\nabla \log p_\lambda = - \nabla f^\lambda -\nabla g $ is always Lipschitz continuous, with $\nabla f^{\lambda} (\kappa) = \big(\kappa - {\rm prox}_f^{\lambda} (\kappa) \big)/\lambda$, where ${\rm prox}_f^{\lambda} (\kappa)$ is the {\it proximity operator} of $f$ at $\kappa$ defined as
\begin{equation} \label{eqn:prox-ope}
{\rm prox}_f^{\lambda} (\kappa)  \equiv \argminT_{\hat{\kappa}\in \mathbb{C}^N}  \left \{ f(\hat{\kappa}) + \|\hat{\kappa} - \kappa\|^2/2\lambda \right \}.
\end{equation}
Replacing $\nabla \log p$ by $\nabla \log p_\lambda$ in the Markov chain of ULA and MALA given in \eqref{eqn:ldp-d} yields,
\begin{equation}\label{eqn:myula-ite}
\begin{split}
{\vect l}^{(m+1)} = & \ \left (1 - \frac{\delta}{\lambda}\right ) {\vect l}^{(m)} + \frac{\delta}{\lambda}  {\rm prox}_{f}^{\lambda} ({\vect l}^{(m)})   - \delta \nabla g({\vect l}^{(m)})\\
			    &  + \sqrt{2\delta} {\vect w}^{(m)},
\end{split}
\end{equation}
which is named the MYULA algorithm (Moreau-Yosida regularised ULA). The MYULA chain \eqref{eqn:myula-ite}, with small $\lambda$, efficiently delivers samples that are approximately distributed according to the posterior $p(\kappa | \gamma)$. By analogy with the process used to obtain MALA from ULA, we create the Px-MALA (proximal MALA) after involving an MH (Metropolis-Hasting) accept-reject step in MYULA.

Essentially, the main difference of the two proximal MCMC methods (MYULA and Px-MALA) is that Px-MALA includes a Metropolis-Hastings step which is used to
correct the bias present in MYULA. Therefore, Px-MALA can provide results with more accuracy, at the expense of a higher computational cost
and slower convergence \citep{[41]}. Note, however, that these MCMC methods (as with any MCMC method) will suffer when scaling to high-dimensional data.
Refer to {\it e.g.} \cite{[40],[41],[11]} for more detailed description of the proximal MCMC methods.

In this article, akin to the experiments performed in \citet{[12]},
we use the proximal MCMC method Px-MALA as a benchmark in the subsequent numerical tests presented in this work.

\section{Approximate Bayesian Uncertainty Quantification} \label{sec:MAPUncertainties}
Though MAP solutions are theoretically optimal (most probable, given the data) one is often interested in the posterior distribution about this MAP point estimate \emdash a necessity if one wishes to be confident in one's result. As described in section \ref{sec:MCMC} we can recover this posterior distribution completely using proximal MCMC techniques such as Px-MALA. However, these approaches are highly computationally demanding. They are feasible in the planar setting at a resolution of $256\times256$, where computation is of $\mathcal{O}(30 \; \text{hours})$, but quickly become unrealistic for high resolutions.
\par
More fundamentally, if we extend mass-mapping from the planar setting to the spherical setting \citep{[3]} the wavelet and measurement operators become more complex \emdash fast Fourier transforms are replaced with full spherical harmonic transforms \emdash and recovery of the posterior \textit{via} MCMC techniques become highly computationally challenging at high resolutions.
\par
In stark contrast to traditional MCMC techniques, recent advances in probability density theory have paved the way for efficient calculation of theoretically conservative approximate Bayesian credible regions of the posterior \citep{[10]}. This approach allows us to extract useful information from the posterior without explicitly having to sample the full posterior. Crucially, this approach is shown to be many orders of magnitude less computationally demanding than \textit{state-of-the-art} MCMC methods \citep{[11]} and can be parallelized and distributed.
\par
In the following section we formally define the concept of a Bayesian credible region of the posterior. We discuss limitations of computing these credible regions and highlight recently proposed approximations to Bayesian credible region.
Finally we outline recently developed computationally efficient uncertainty quantification techniques which can easily scale to high-dimensional data. Specifically, we introduce the concept of \textit{local credible intervals} (\textit{cf.} pixel level error bars) presented first in \citet{[12]} to the weak lensing setting.

\subsection{Highest Posterior Density}
A posterior credible region at $100(1-\alpha)\%$ confidence is a set $C_{\alpha} \in \mathbb{C}^N$ which satisfies
\begin{equation} \label{eq:CredibleIntegral}
p(\kappa \in C_{\alpha}|\gamma) = \int_{\kappa \in \mathbb{C}^N} p(\kappa|\gamma)\mathbb{I}_{C_{\alpha}}d\kappa = 1 - \alpha.
\end{equation}
Generally there are many regions which satisfy this constraint. The minimum volume, and thus decision-theoretical optimal \citep{[19]}, region is the \textit{highest posterior density} (HPD) credible region, defined to be
\begin{equation}
C_{\alpha} := \lbrace \kappa : f(\kappa) + g(\kappa) \leq \epsilon_{\alpha} \rbrace,
\end{equation}
where $f(\kappa)$ is the prior and $g(\kappa)$ is the data fidelity (likelihood) term. In the above equation $\epsilon_{\alpha}$ is an isocontour (\textit{i.e.} level-set) of the log-posterior set such that the integral constraint in equation (\ref{eq:CredibleIntegral}) is satisfied. In practice the dimension $N$ of the problem is large and the calculation of the true HPD credible region is difficult to compute.
\par
Recently a conservative approximation of $C_{\alpha}$ has been derived \citep{[10]}, which can be used to tightly constrain the  HPD credible region without having to explicitly calculate the integral in equation (\ref{eq:CredibleIntegral}):
\begin{equation}
C^{\prime}_{\alpha} := \lbrace \kappa : f(\kappa) + g(\kappa) \leq \epsilon^{\prime}_{\alpha} \rbrace.
\end{equation}
By construction this approximate credible-region is conservative, which is to say that $\lbrace C_{\alpha} \subset C_{\alpha}^{\prime} \rbrace$. Importantly, this means that if a $\kappa$ map does \textbf{not} belong to $C_{\alpha}^{\prime}$ then it necessarily \textbf{cannot} belong to $C_{\alpha}$. The approximate level-set threshold $\epsilon^{\prime}_{\alpha}$ at confidence $100(1-\alpha)\%$ is given by
\begin{equation}
\epsilon^{\prime}_{\alpha} = f(\kappa^{\text{map}}) + g(\kappa^{\text{map}}) + \tau_{\alpha} \sqrt{N} + N,
\end{equation}
where we recall $N$ is the dimension of $\kappa$. The constant $\tau_{\alpha} = \sqrt{16 \log(3 / \alpha)}$ quantifies the envelope required such that the HPD credible-region is a sub-set of the approximate HPD credible-region. There exists an upper bound on the error introduced through this approximation, which is given by
\begin{equation} \label{eq:ErrorLevelSet}
0 \leq \epsilon^{\prime}_{\alpha} - \epsilon_{\alpha} \leq \eta_{\alpha} \sqrt{N} + N,
\end{equation}
where the factor $\eta_{\alpha} = \sqrt{16 \log (3/\alpha)} + \sqrt{1/\alpha}$. This approximation error scales at most linearly with $N$. As will be shown in this paper this upper bound is typically extremely conservative in practice, and the error small.
\par
We now introduce a recently proposed strategy for uncertainty quantification building on the concept of approximate HPD credible-regions. For further details on the strategy we recommend the reader see related work \citep{[12]}.

\subsection{Local Credible Intervals}
Local credible intervals can be interpreted as error bars on individual pixels or super-pixel regions (collection of pixels) of a reconstructed $\kappa$ map. This concept can be applied to any method for which the HPD credible-region (and thus the approximate HPD credible-region) can be computed. Mathematically local credible intervals can be computed as follows \citep{[12]}.
\par
Select a partition of the $\kappa$ domain $\Omega = \cup_i \Omega_i$ such that super-pixels $\Omega_i$ (\textit{e.g.} an $8 \times 8$ block of pixels) are independent sub-sets of the $\kappa$ domain $\Omega_i \cap \Omega_j \: = \varnothing, \: \forall \: \lbrace i \neq j \rbrace$. Clearly, provided the super-pixels $\Omega_i$ completely tessellate $\Omega$ they can be of arbitrary dimension. We define indexing notation on the super-pixels $\Omega_i$ via the index operator $\zeta_{\Omega_i}$ which satisfy analogous relations to the standard set indicator function given in equation (\ref{eq:indicator}) \emdash \textit{i.e.} $\zeta_{\Omega_i} = 1$ if the pixel of the convervence map $\kappa$ belongs to $\Omega_i$ and $0$ otherwise.
\par
For a given super-pixel region $\Omega_i$ we quantify the uncertainty by finding the upper and lower bounds $\xi_{+,\Omega_i}$,  $\xi_{-,\Omega_i}$ respectively, which raise the objective function above the approximate level-set threshold $\epsilon^{\prime}_{\alpha}$ (or colloquially, `saturate the HPD credible region $C_{\alpha}^{\prime}$'). In a mathematical sense these bounds are defined by
\begin{equation}
\xi_{+,\Omega_i} = \maxT_{\xi} \big \lbrace \xi | f(\mathbf{\kappa}_{i,\xi}) + g(\mathbf{\kappa}_{i,\xi}) \leq \epsilon^{\prime}_{\alpha}, \: \forall \xi \in \mathbb{R} \big \rbrace
\end{equation}
and
\begin{equation}
\xi_{-,\Omega_i} = \minT_{\xi} \big \lbrace \xi | f(\mathbf{\kappa}_{i,\xi}) + g(\mathbf{\kappa}_{i,\xi}) \leq \epsilon^{\prime}_{\alpha}, \: \forall \xi \in \mathbb{R} \big \rbrace,
\end{equation}
where $\mathbf{\kappa}_{i,\xi} = \kappa^{\text{map}} (\mathbf{I} - \zeta_{\Omega_i}) + \xi \zeta_{\Omega_i}$ is a surrogate solution where the super-pixel region has been replaced by a uniform intensity $\xi$. We then construct the difference image $\sum_i(\xi_{+,\Omega_i} - \xi_{-,\Omega_i})$ which represents the length of the local credible intervals (\textit{cf.} error bars) on given super-pixel regions at a confidence of $100(1-\alpha)\%$.
\par
In this paper we locate $\xi_{\pm}$ iteratively \textit{via} bisection, though faster converging algorithms could be used to further increase computational efficiency. A schematic diagram for constructing local credible intervals is found in Figure \ref{fig:LocalCredibleIntervals}. Conceptually, this is finding the \textbf{maximum} and \textbf{minimum} constant values which a super-pixel region could take, at $100(1-\alpha)\%$ confidence \emdash which is effectively Bayesian error bars on the convergence map.
\par
In plain english, starting from the MAP convergence solution $\kappa^{\text{map}}$ \emdash at which all pixels are in positions which minimize the objective function \emdash we then select a sub-set of the pixels (\textit{e.g.} an $8 \times 8$ block of pixels). We start by averaging the pixels in the block which is selected. We then set the pixels within this block to the average value. Following this we iteratively raise/lower the now uniform value of the pixels within this block whilst keeping the rest of the image fixed. After each iteration we check if the surrogate solution ($\kappa^{\text{map}}$ with the block of interest replaced by some constant value) is an acceptable solution (\textit{i.e.} the objective function is below the threshold $\epsilon_{\alpha}^{\prime}$). We find the values (upper and lower bounds) at which objective function is equal to the threshold $\epsilon_{\alpha}^{\prime}$. We then take the difference between these bounds, which is the \textit{local credible interval} for a given `block of interest' (super-pixel region).
\par

%%%%%%%%%%%%%%%%%%%%%%%%%%%%%%%%%%%%%%%%%%%%%%%%%%%%%%%%%%%%%%%%%%%%%%%%%%%%%%%%%%%%%%%%%%
\begin{figure}
\begin{center}
\begin{tikzpicture}[node distance = 2cm, auto]
    % Place nodes
    \node [block, text width=8em] (0) {Calculate MAP solution: $\kappa^{\text{map}}$};
    \node [block, below of=0, node distance=1.6cm, text width=8em] (1) {Define: super-pixel $\Omega_i$};
    \node [block, below of=1, node distance=1.6cm, text width=8em] (2) {Calculate average: $\xi = \langle \kappa^{\text{map}} \zeta_{\Omega_i} \rangle$};
    \node [block, below of=2, node distance=1.6cm] (3) {Create Surrogate: $\mathbf{\kappa}_{i,\xi} = \kappa^{\text{map}} (\mathbf{I} - \zeta_{\Omega_i}) + \xi \zeta_{\Omega_i}$};
    \node [decision, below of=3, node distance=2.0cm] (4) { $\mathbf{\kappa}_{i,\xi} \in C_{\alpha}^{\prime}$ };
    \node [block, right of=3, node distance=4.0cm, text width=8em] (5) {$\xi \leftarrow \xi \; \pm$ step-size};
    \node [block, below of=4, node distance=2.0cm] (6) {Max/Min: $\xi_{\pm} = \xi$};
    % Draw edges
    \path [line] (0) -- (1);
    \path [line] (1) -- (2);
    \path [line] (2) -- (3);
    \path [line] (3) -- (4);
    \path [line,dashed] (4) -| node[near start] {Yes}(5);
    \path [line,dashed] (5) -- (3);
    \path [line,dashed] (4) -- node {No}(6);

\end{tikzpicture}
\caption{Schematic of the process to construct local credible intervals. At each iterative step the super-pixel region is uniformly increased (decreased) by a step-size. Once the level-set threshold $\epsilon_{\alpha}^{\prime}$ is saturated the iteration is terminated. Note that this diagram does not represent the bisection method that is adopted in this article, which is a little more involved, but just a simple iterative scheme for conceptualization.}
\label{fig:LocalCredibleIntervals}
\end{center}
\end{figure}
%%%%%%%%%%%%%%%%%%%%%%%%%%%%%%%%%%%%%%%%%%%%%%%%%%%%%%%%%%%%%%%%%%%%%%%%%%%%%%%%%%%%%%%%%%

%%%%%%%%%%%%%%%%%%%%%%%%%%%%%%%%%%%%%%%%%%%%%%%%%%%%%%%%%%%%%%%%
% Bolshoi 7 8 combined
\begin{figure*}
	\centering
	\includegraphics[width=\textwidth]{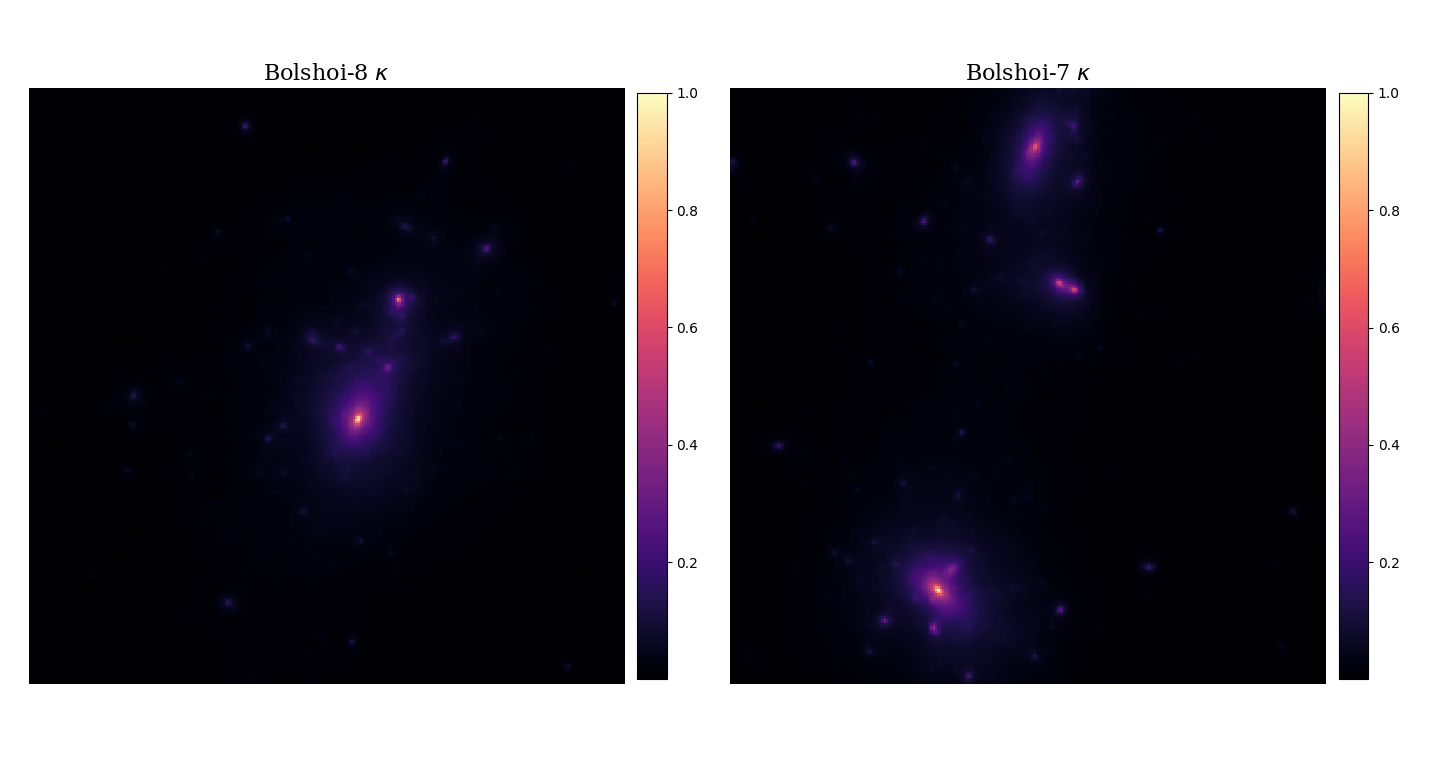}
    \caption{Two of the largest clusters extracted from the Bolshoi simulation database, labeled as Bolshoi 7 and 8 somewhat arbitrarily. In both cases at least one massive sub-halo is located within the FoF (friends of friends) sub-catalog, as can be clearly seen.}
    \label{fig:Bolshoi_7_8_combined}
\end{figure*}
%%%%%%%%%%%%%%%%%%%%%%%%%%%%%%%%%%%%%%%%%%%%%%%%%%%%%%%%%%%%%%%%

%%%%%%%%%%%%%%%%%%%%%%%%%%%%%%%%%%%%%%%%%%%%%%%%%%%%%%%%%%%%%%%%
% Buzzard 1 2 combined
\begin{figure*}
	\centering
	\includegraphics[width=\textwidth]{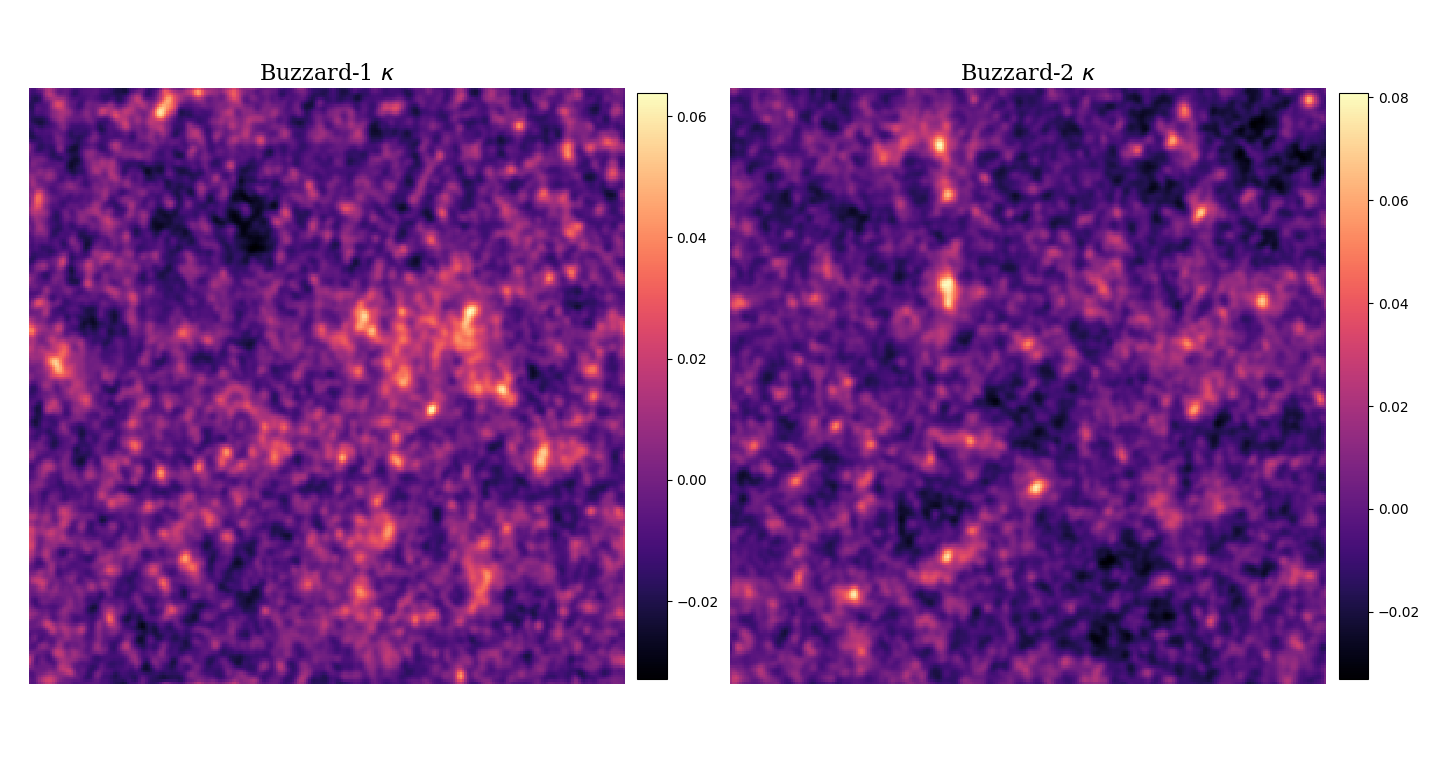}
    \caption{Two $\sim 1.2 \deg^{2}$ planar random extractions from the Buzzard V-1.6 N-body simulation catalog, each containing $\mathcal{O}(10^6)$ galaxies.}
    \label{fig:Buzzard_1_2_combined}
\end{figure*}
%%%%%%%%%%%%%%%%%%%%%%%%%%%%%%%%%%%%%%%%%%%%%%%%%%%%%%%%%%%%%%%%

\section{Evaluation on Simulations} \label{sec:Testing}
For computing Bayesian inference problems one would ideally adopt an MCMC approach as they are (assuming convergence) guaranteed to produce optimal results, however these approaches are computationally demanding and can often be computationally infeasible. Therefore it is beneficial to adopt approximate but significantly computationally cheaper methods, such as the MAP estimation approach reviewed in this article \emdash first presented in \citet{[M1]}.
\par
However, the approximation error introduced through these approximate methods must be ascertained. Therefore, in this section we benchmark the uncertainties reconstructed \textit{via} our MAP algorithm to those recovered by the \textit{state-of-the-art} proximal MCMC algorithm, Px-MALA \citep{[40],[41]}. Additionally we compare the computational efficiencies of both approaches, highlighting the computational advantages provided by approximate methods.
\par
For simplicity and brevity throughout we will refer to any uncertainties recovered \textit{via} our
aforementioned \textit{maximum a posteriori} reconstruction method as `MAP uncertainties'. Additionally we will
refer to the \textit{maximum a posteriori} reconstruction method discussed throughout this paper as
`MAP algorithm`.

\subsection{Datasets} \label{sec:Data}
We select four test convergence fields: two large scale Buzzard N-body simulation \citep{DeRose2018,wechsler2018} planar patches selected at random; and two of the largest dark matter halos from the Bolshoi N-body simulation \citep{[20]}. This selection is chosen such as to provide illustrative examples of the uncertainty quantification techniques in both cluster and wider-field weak lensing settings.

\subsubsection{Bolshoi N-body}
The Bolshoi cluster convergence maps used were produced from 2 of the largest halos in the Bolshoi N-body simulation. These cluster were selected for their large total mass and the complexity of their substructure, as can be seen in Figure \ref{fig:Bolshoi_7_8_combined}.
\par
Raw particle data was extracted from the Bolshoi simulation using CosmoSim\footnote{https://www.cosmosim.org}, and was then gridded into $1024\times1024$ images. These images inherently contain shot-noise and so were passed through a multi-scale Poisson denoising algorithm before being re-gridded to $256\times256$.
\par
The denoising algorithm consisted of a forward Anscombe transform (to Gaussianise the noise), several TV-norm (total-variation) denoising optimizations of different scale, before finally inverse Anscombe transforming. Finally, the images were re-scaled onto $[0,1]$ \emdash a similar denoising approach for Bolshoi N-body simulations was adopted in related articles \citet{[6]}.

\subsubsection{Buzzard N-body}
The Buzzard v-1.6 shear catalogs are extracted by ray-tracing from a full end-to-end N-body simulation. The origin for tracing is positioned in the corner of the simulation box and so the catalog has $25\%$ sky coverage. Access to the Buzzard simulation catalogs was provided by the LSST-DESC collaboration\footnote{http://lsst-desc.org}.
\par
In the context of this paper we restrict ourselves to working on the plane, and as such we extracted smaller planar patches. To do so we first project the shear catalog into a coarse HEALPix\footnote{http://healpix.sourceforge.net/documentation.php}\citep{Gorski2005HEALPix} griding (with $N_{\text{side}}$ of 16). Inside each HEALPix pixel we tessellate the largest possible square region, onto which we rotate and project the shear catalog. Here HEALPix pixelisation is solely used for its equal area pixel properties.
\par
After following the above procedure, the Buzzard v-1.6 shear catalog reduces to $\sim 3 \times 10^3$ planar patches of angular size $\sim 1.2 \deg^{2}$, with $\sim 4 \times 10^6$ galaxies per patch. In previous work \citep{[M1]} we utilized 60 of these realisations, but for the purpose of this paper we select at random two planar regions to study, which we grid at a $256\times256$ resolution. These plots can be seen in Figure \ref{fig:Buzzard_1_2_combined}.

\subsection{Methodology}
To draw comparisons between our MAP uncertainties and those recovered \textit{via} Px-MALA we conduct the following set of tests on the aforementioned datasets (see section \ref{sec:Data}).
\par
Initially we transform the ground truth convergence $\kappa^{\text{in}}$ into a clean shear field $\gamma^{\text{in}}$ by
\begin{equation}
\gamma^{\text{in}} = \bm{\Phi} \kappa^{\text{in}}.
\end{equation}
This clean set of shear measurements is then contaminated with a noise term $n$ to produce mock noisy observations $\gamma$ such that
\begin{equation}
\gamma = \gamma^{\text{in}} + n.
\end{equation}
For simplicity we choose the noise to be zero mean i.i.d. Gaussian noise of variance $\sigma_n^2$ \emdash \textit{i.e.} $n \sim \mathcal{N}(0,\sigma_n^2)$.
In this setting $\sigma_n$ is calculated such that the signal to noise ratio (SNR) is 20 dB (decibels) where
\begin{equation}
\sigma_n = \sqrt{\frac{\norm{\bm{\Phi} \kappa}_2^2}{N}} \times 10^{-\frac{\text{SNR}}{20}}.
\end{equation}
Throughout this uncertainty benchmarking we use a fiducial noise level of 20 dB. For further details on how a noise level in dB maps to quantities such as galaxy number density and pixel size see \citet{[M1]}. In particular, we draw the reader's attention to \citet[Table C1][]{[M1]}. The noise level of 20 dB considered here is somewhat optimistic (corresponding to between 30 and 100 galaxies per square arcmin for a band-limit of $\sim$400), which is appropriate for the purposes of benchmarking against MCMC simulations, which is the focus of the current article (less optimistic simulations would simply increase the absolute level of the quantified uncertainties but not their relative level).
\par
We then apply our entire reconstruction pipeline \citep{[M1]}, as briefly outlined in section \ref{sec:BayesianInference}, to recover $\kappa^{\text{map}}$, along with the objective function \emdash with regularization parameter $\mu$ and noise variance $\sigma_n^2$. Using these quantities, and the Bayesian framework outlined in sections \ref{sec:HierarchicalBayesianInference} and \ref{sec:MAPUncertainties}, we conduct uncertainty quantifications on $\kappa^{\text{map}}$.
\par
To benchmark the MAP reconstructed uncertainties we first construct an array of \textit{local credible interval} maps described in section \ref{sec:MAPUncertainties} for super-pixel regions of sizes $[4, 8, 16]$ at $99\%$ confidence. These local credible interval maps are then compared to those recovered from the full MCMC analysis of the posterior.
\par
We adopt two basic statistical measures to compare each set of recovered local credible interval maps: the Pearson correlation coefficient $r$; and the recovered SNR. The Pearson correlation coefficient between our MAP local credible interval map $\xi^{\text{map}} \in \mathbb{R}^{N^{\prime}}$ and the Px-MALA local credible interval map $\xi^{\text{px}} \in \mathbb{R}^{N^{\prime}}$, where $N^{\prime}$ is the dimension of the super-pixel space, is defined to be
\begin{equation}
r = \frac{ \sum_{i=1}^{N^{\prime}} ( \xi^{\text{map}}(i) - \bar{\xi}^{\text{map}} ) ( \xi^{\text{px}}(i) - \bar{\xi}^{\text{px}} ) }{ \sqrt{\sum_{i=1}^{N^{\prime}} ( \xi^{\text{map}}(i) - \bar{\xi}^{\text{map}} )^2} \sqrt{\sum_{i=1}^{N^{\prime}} ( \xi^{\text{px}}(i) - \bar{\xi}^{\text{px}} )^2}  },
\end{equation}
where $\bar{x} = \langle x \rangle$. The correlation coefficient $r \in [-1,1]$ quantifies the structural similarity between two datasets: 1 indicates maximally positive correlation, 0 indicates no correlation, and -1 indicates maximally negative correlation.
\par
The second of our two statistics is the recovered SNR which is calculated between $\xi^{\text{map}}$ and $\xi^{\text{px}}$ to be
\begin{equation}
\text{SNR} = 20 \times \log_{10}\Bigg (\frac{\norm{\xi^{\text{px}}}_2}{\norm{\xi^{\text{px}} - \xi^{\text{map}}}_2} \Bigg ),
\end{equation}
where $\xi^{\text{px}}$ recovered by Px-MALA is assumed to represent the ground truth Bayesian local credible interval, and $\norm{.}_2$ is the $\ell_2$-norm. The SNR is a measure of the absolute similarity of two maps \emdash in this context, rather than the structural correlation which is encoded into $r$, the SNR is a proxy measure of the relative magnitudes of the two datasets. Additionally, we compute the root mean squared percent error (RMSE),
\begin{equation}
\text{RMSE} = 100  \times \Bigg (\frac{\norm{\xi^{\text{px}} - \xi^{\text{map}}}_2}{\norm{\xi^{\text{px}}}_2} \Bigg ) \%.
\end{equation}
\par
Conceptually, the SNR roughly compares the absolute magnitudes of recovered local credible intervals and the Pearson correlation coefficient gives a rough measure of how geometrically similar the local credible intervals are. In this sense the closer $r$ is to 1 the more similar the recovered local credible intervals are, and the higher the SNR the smaller the approximation error given by equation (\ref{eq:ErrorLevelSet}). Thus, a positive result is quantified by both large correlation and large SNR.

\subsection{Results}
\begin{figure*}
	\centering
	\includegraphics[width=\textwidth]{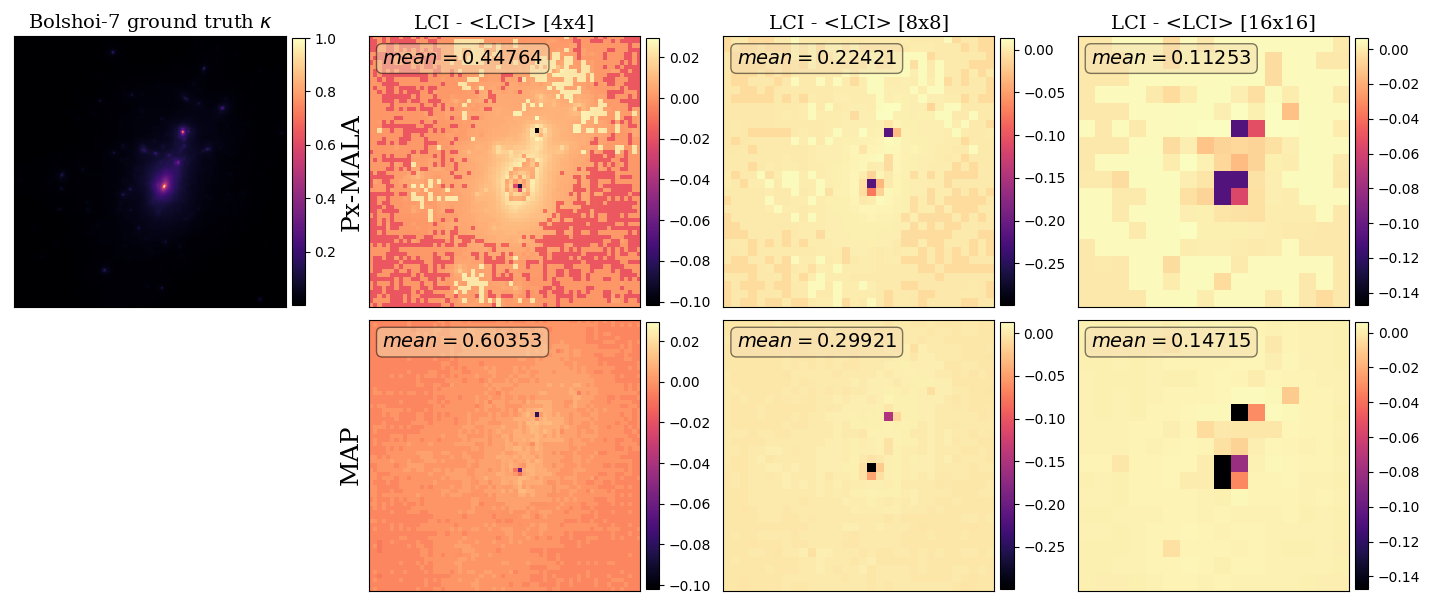}
    \includegraphics[width=\textwidth]{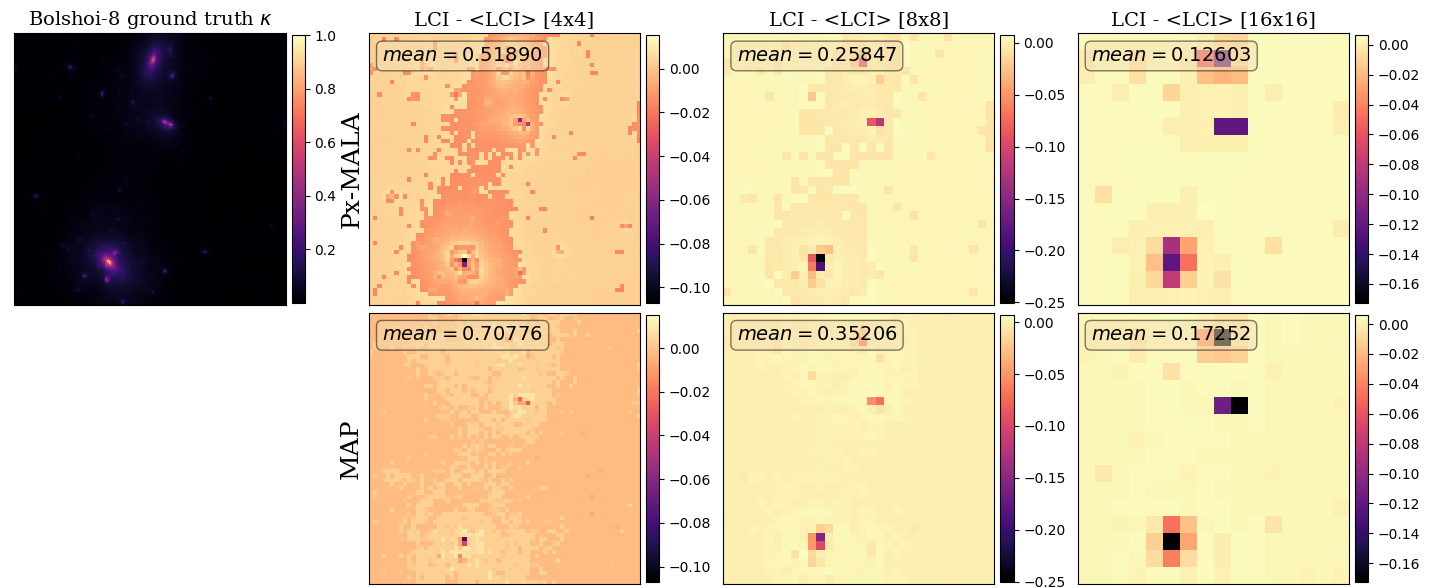}
    \caption{Local Credible Intervals (\textit{cf.\ Bayesian error bars}) at $99\%$ confidence for the Bolshoi-7 (\textbf{top}) and Bolshoi-8 (\textbf{bottom}) cluster sparse reconstruction in both the Px-MALA setting (\textit{top}) and MAP (\textit{bottom}) for super-pixel regions of dimension $(4\times4), (8\times8),$ and $(16\times16)$ \emdash left to right respectively. Note that these plots display the variation about the mean of each set of LCI's, with the mean being given numerically in the sub-figure legends \emdash this is done to best display the topological similarity whilst also conveying the absolute difference in size between the methods. Notice that the mean of the MAP LCIs is in all cases larger than that of the corresponding Px-MALA LCIs.} Further note that the smaller the dimension of the super-pixel the larger the local credible interval which is because adjusting fewer pixels raises the objective function by less, and so the smaller super-pixels can be raised/lowered by more before saturating the level-set threshold. All numerical results are displayed in Table \ref{tab:data}.
    \label{fig:LCI_Bolshoi_7_comparison}
\end{figure*}

\begin{figure*}
	\centering
	\includegraphics[width=\textwidth]{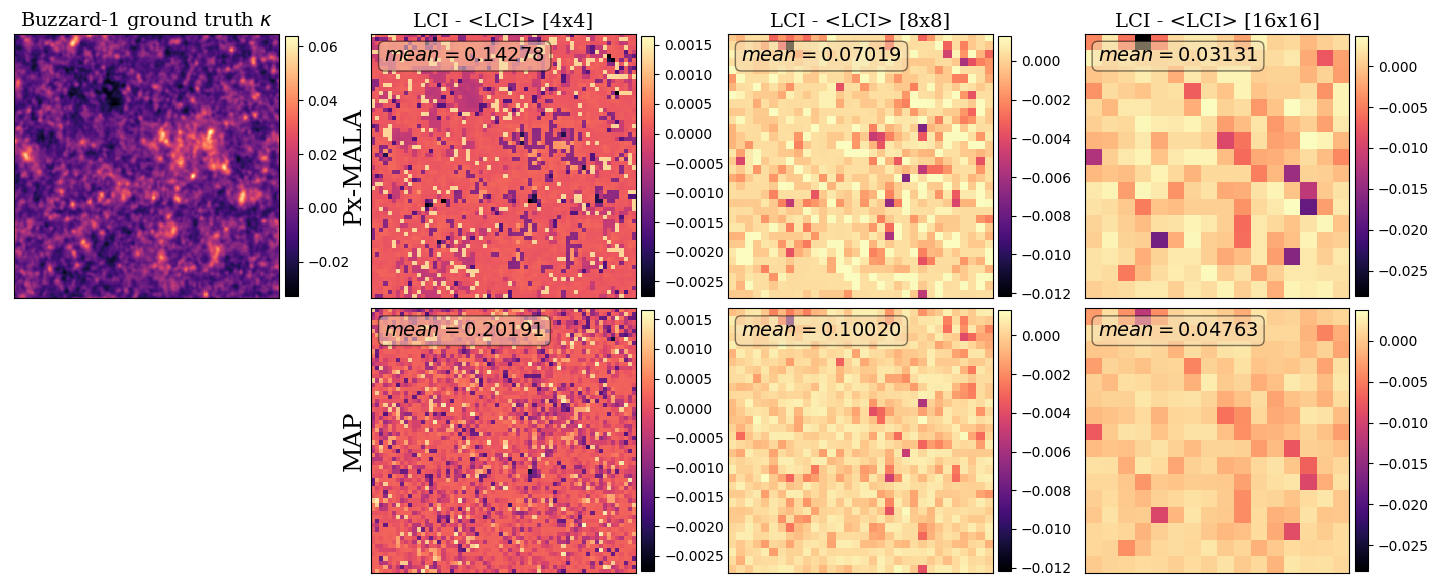}
    \includegraphics[width=\textwidth]{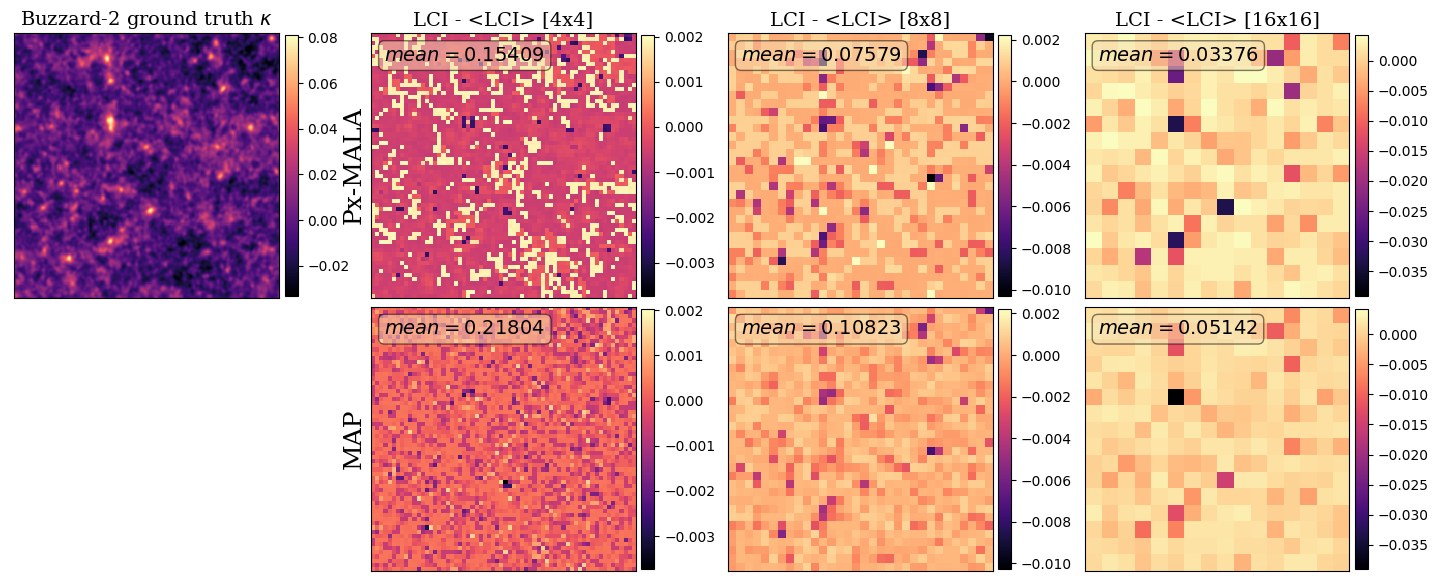}
    \caption{Local Credible Intervals (\textit{cf.\ Bayesian error bars}) at $99\%$ confidence for the Buzzard-1 (\textbf{top}) and Buzzard-2 (\textbf{bottom}) cluster sparse reconstruction in both the Px-MALA setting (\textit{top}) and MAP (\textit{bottom}) for super-pixel regions of dimension $(4\times4), (8\times8),$ and $(16\times16)$ \emdash left to right respectively. Note that these plots display the variation about the mean of each set of LCI's, with the mean being given numerically in the sub-figure legends \emdash this is done to best display the topological similarity whilst also conveying the absolute difference in size between the methods. Notice that the mean of the MAP LCIs is in all cases larger than that of the corresponding Px-MALA LCIs.} Further note that the smaller the dimension of the super-pixel the larger the local credible interval which is because adjusting fewer pixels raises the objective function by less, and so the smaller super-pixels can be raised/lowered by more before saturating the level-set threshold. All numerical results are displayed in Table \ref{tab:data}.
    \label{fig:LCI_Buzzard_1_comparison}
\end{figure*}

As can be seen in Figures \ref{fig:LCI_Bolshoi_7_comparison} and \ref{fig:LCI_Buzzard_1_comparison} the local credible intervals recovered through our sparse hierarchical Bayesian formalism are at all times larger than those recovered via Px-MALA \emdash confirming that the uncertainties are conservative, as proposed in section \ref{sec:MAPUncertainties}. Moreover, a strong correlation between the reconstructions can be seen.

%%%%%%%%%%%%%%%%%%%%%%%%%%%%%%%%%%%%%%%%%%%%%%%%%%%%%%%%%%%%%%%%
% Local Credible Interval Bolshoi 7
%%%%%%%%%%%%%%%%%%%%%%%%%%%%%%%%%%%%%%%%%%%%%%%%%%%%%%%%%%%%%%%%

%%%%%%%%%%%%%%%%%%%%%%%%%%%%%%%%%%%%%%%%%%%%%%%%%%%%%%%%%%%%%%%%
% Local Credible Interval Buzzard 1
%%%%%%%%%%%%%%%%%%%%%%%%%%%%%%%%%%%%%%%%%%%%%%%%%%%%%%%%%%%%%%%%

%%%%%%%%%%%%%%%%%%%%%%%%%%%%%%%%%%%%%%%%%%%%%%%%%%%%%%%%%%%%%%%%
% Buzzard 1 LCI data-table
\begin{table}
	\centering
	\caption{Comparisons between the local credible interval maps recovered \textit{via} MAP and those recovered \textit{via} Px-MALA. Note that higher super-pixels corresponds to coarser resolutions whereas smaller super-pixels leads to higher resolution. This is because the super-pixel size is the size of the groups of pixels used to tile the original image \emdash therefore larger tiling components leads to fewer tiles, and therefore lower resolution.}
	\label{tab:data}
	\begin{tabular}{l|c|c|c|r} % four columns, alignment for each
		\hline
        \hline
		\textbf{Super} & \textbf{Pearson} & \textbf{SNR} & \textbf{RMSE}  \\
        \textbf{Pixel} & \textbf{Correlation} & \textbf{(dB)} & \textbf{Error}\\
		\hline
        \multicolumn{4}{|c|}{\textbf{Bolshoi-7}}\\
        \hline
		4x4 & 0.463 & 11.737 & 25.892 $\%$\\
		8x8 & 0.848 & 11.994 & 25.137 $\%$\\
		16x16 & 0.945 & 12.509 & 23.690 $\%$\\
%         32x32 & 0.937917 & 11.6928 & 26.0233 $\%$&\checkmark \\
		\hline
        \multicolumn{4}{|c|}{\textbf{Bolshoi-8}}\\
        \hline
		4x4 & -0.168 & 11.467 & 26.710 $\%$\\
		8x8 & 0.929 & 11.490 & 26.637 $\%$\\
		16x16 & 0.941 & 11.350 & 27.070  $\%$\\
%         32x32 & 0.921363 & 10.2934 & 30.5724 $\%$ & \checkmark \\
        \hline
        \multicolumn{4}{|c|}{\textbf{Buzzard-1}}\\
        \hline
		4x4 & 0.164 & 10.666 & 29.289 $\%$\\
		8x8 & 0.916 & 10.473 & 29.948 $\%$\\
		16x16 & 0.984 & 9.262 & 34.427 $\%$\\
%         32x32 & 0.633823 & 2.34941 & 76.3009 $\%$& \checkmark \\
		\hline
        \multicolumn{4}{|c|}{\textbf{Buzzard-2}}\\
        \hline
		4x4 & 0.140 & 10.653 & 29.333 $\%$\\
		8x8 & 0.904 & 10.465 & 29.973 $\%$\\
		16x16 & 0.926 & 9.217 & 34.605 $\%$\\
%         32x32 & 0.643482 & 2.60863 & 74.0574 $\%$ & \checkmark \\
		\hline
        \hline
	\end{tabular}
\end{table}
%%%%%%%%%%%%%%%%%%%%%%%%%%%%%%%%%%%%%%%%%%%%%%%%%%%%%%%%%%%%%%%%
\par
The largest correlation coefficients $r$ are observed for super-pixel regions of dimension $16\times16$ in all cases ($\langle r \rangle \approx 0.9$), peaking as high as 0.98 for the Buzzard 1 extraction \emdash which constitutes a near maximal correlation, and thus an outstanding topological match between the two recovered local credible intervals.
\par
Additionally, in the majority of cases the recovered SNR is $\geq 10$ dB \emdash in some situations rising as high as $\approx 13$ dB (corresponding to $\approx 20 \%$ RMSE percent error) \emdash which indicates that the recovered MAP uncertainties are close in magnitude to those recovered \textit{via} Px-MALA.
\par
However, for super-pixels with dimension $4\times4$ the structural correlation between $\xi^{\text{map}}$ and $\xi^{\text{px}}$ becomes small \emdash in one case becoming marginally negatively correlated. This is likely to be a direct result of the error given by equation (\ref{eq:ErrorLevelSet}) inherited from the definition of the approximate HPD credible region \emdash as this approximation has the side-effect of smoothing the posterior hyper-volume, and for small super-pixels the hyper-volume is typically not smooth, thus the correlation coefficient $r$ decreases.
\par
We conducted additional tests for large $32 \times 32$ dimension super-pixels, which revealed a second feature of note.  For particularly large super-pixel regions ($32\times32$ or larger) the SNR becomes small for both Buzzard maps. This is a result of the assumption that within a super-pixel there exists a stable mean which is roughly uniform across the super-pixel. Clearly, for buzzard type data, on large scales this breaks down and so the recovered local credible intervals deviate from those recovered \textit{via} Px-MALA. It is important to stress this is a breakdown of the assumptions made when constructing local credible intervals and not an error of the approximate HPD credible region.
\par
The numerical results are summarised in Table \ref{tab:data}. Typically, structures of interest in recovered convergence maps cover super-pixel regions of roughly $8\times8$ to $16\times16$, and so for most realistic applications our MAP uncertainties match very well with those recovered through Px-MALA. In most situations weak lensing data is gridded such that it best represents the features of interest, and so structures of interest (by construction) typically fall within $8\times8$ to $16\times16$ dimension super-pixel regions for $256 \times 256$ gridded images \emdash for higher resolution images the structures of interest, and corresponding optimal super-pixels will follow a similar ratio.
\par
Overall, we find a very close relation between the local credible intervals recovered through our MAP algorithm with those recovered \textit{via} Px-MALA \emdash a state-of-the-art MCMC algorithm. We find that MAP and Px-MALA local credible intervals are typically strongly topologically correlated (pearson correlation coefficient $\approx 0.9$) in addition to being physically tight (RMSE error of $\approx 20-30 \%$). Moreover, we find that the MAP local credible intervals are, everywhere, larger than the Px-MALA local credible intervals, corroborating the assertion that the approximate HPD level-set threshold $\epsilon_{\alpha}^{\prime}$ is in fact conservative.

%%%%%%%%%%%%%%%%%%%%%%%%%%%%%%%%%%%%%%%%%%%%%%%%%%%%%%%%%%%%%%%%
% Buzzard 1 LCI data-table
\begin{table}
	\centering
	\caption{Numerical comparison of computational time of Px-MALA and MAP. The MAP approach typically takes $\mathcal{O}(10^{-1})$ seconds, compared to Px-MALA's $\mathcal{O}(10^5)$ seconds. Therefore for linear reconstructions MAP is close to $\mathcal{O}(10^6)$ times faster.}
	\label{tab:data_computational_posterior}
	\begin{tabular}{l|c|c|c|r} % four columns, alignment for each
		\hline
        \hline
		\textbf{Px-MALA} & \textbf{MAP} & \multirow{2}{*}{\textbf{Ratio}} \\
        \textbf{Time (s)} & \textbf{Time (s)} & \\
		\hline
        \hline
        & \textbf{Buzzard-1}\\
        \hline
		133761 & 0.182 & 0.734 $\times 10^6$ \\
		\hline
        & \textbf{Buzzard-2}\\
        \hline
		141857 & 0.175 & 0.811 $\times 10^6$\\
		\hline
        & \textbf{Bolshoi-7}\\
        \hline
		95339 & 0.153 & 0.623 $\times 10^6$\\
		\hline
        & \textbf{Bolshoi-8}\\
        \hline
		92929 & 0.143 & 0.650 $\times 10^6$\\
		\hline
        \hline
	\end{tabular}
\end{table}
%%%%%%%%%%%%%%%%%%%%%%%%%%%%%%%%%%%%%%%%%%%%%%%%%%%%%%%%%%%%%%%%

We now compare the computational efficiency of our sparse Bayesian reconstruction algorithm against Px-MALA. It is worth noting that all Px-MALA computation was done on a high performance workstation (with 24 CPU cores and 256Gb of memory), whereas all MAP reconstructions were done on a standard 2016 MacBook Air.
The computation time for MAP estimation is found to be $\mathcal{O}$(seconds) whereas the computation time for Px-MALA is found to be $\mathcal{O}$(days). Specifically, we find the MAP reconstruction algorithm is of $\mathcal{O}(10^6)$ (typically $\geq 8 \times 10^5$) times faster than the \textit{state-of-the-art} Px-MALA MCMC algorithm. Moreover, the MAP reconstruction algorithm supports algorithmic structures that can be highly parallelized and distributed.

\section{Conclusions} \label{sec:Conclusions}
In this article we introduce the concept of local credible intervals (\textit{cf.} pixel-level error bars) \emdash  developed in previous work and applied in the radio-interferometric setting \emdash to the weak lensing setting as an additional form of uncertainty quantification. Utilizing local credible intervals we validate the sparse hierarchical Bayesian mass-mapping formalism presented in previous work \citep{[M1]}. Specifically we compare the local credible intervals recovered \textit{via} the MAP formalism and those recovered \textit{via} a complete MCMC analysis \emdash from which the true posterior is effectively recovered.
\par
To compute the asymptotically exact posterior we utilize Px-MALA \emdash a \textit{state-of-the-art} proximal MCMC algorithm. Using the local credible intervals; we benchmark the MAP uncertainty reconstructions against Px-MALA.
\par
Quantitatively, we compute the Pearson correlation coefficient ($r$, as a measure of the correlation between hyper-volume topologies), recovered signal to noise ratio and the root mean squared percentage error (SNR and RMSE, both as measures of how tightly constrained is the absolute error).
\par
We find that for a range of super-pixel dimensions the MAP and Px-MALA uncertainties are strongly topologically correlated ($r \geq 0.9$). Moreover, we find the RMSE to typically be $\sim 20-30 \%$ which is tightly constrained when one considers this is a conservative approximation along each of at least $\mathcal{O}(10^{3})$ dimensions.
\par
Additionally we compare the computational efficiency of Px-MALA and our MAP approach. In a $256\times256$ setting, the computation time of the  MAP approach was $\mathcal{O}$(seconds) whereas the compuattion time for Px-MALA was $\mathcal{O}$(days). Overall, the MAP approach is shown to be $\mathcal{O}(10^6)$ times faster than the \textit{state-of-the-art} Px-MALA algorithm.
\par
A natural progression is to extend the planar sparse Bayesian algorithm to the sphere, which will be the aim of upcoming work \emdash a necessity when dealing with wide-field stage \rom{4} surveys such as LSST\footnote{https://www.lsst.org} and EUCLID\footnote{http://euclid-ec.org}. Additionally, we will expand the set of uncertainty quantification techniques to help propagate principled Bayesian uncertainties into the set of higher-order statistics typically computed on the convergence field.

\section*{Acknowledgements}

Author contributions are summarised as follows.
MAP: methodology, data curation, investigation, software, visualisation, writing - original draft;
XC: methodology, investigation, software, writing - review \& editing;
JDM: conceptualisation, methodology, project administration, supervision, writing - review \& editing;
MP: methodology, software, writing - review \& editing;
TDK: methodology, supervision, writing - review \& editing.

This paper has undergone internal review in the LSST Dark Energy Science Collaboration. The internal reviewers were Chihway  Chang, Tim Eifler, and Francois Lanusse.
The author thank the development teams of SOPT. MAP is supported by the Science and Technology Facilities Council (STFC). TDK is supported by a Royal Society University Research Fellowship (URF). This work was also supported by the Engineering and Physical Sciences Research Council (EPSRC) through grant EP/M011089/1 and by the Leverhulme Trust.
The DESC acknowledges ongoing support from the Institut National de Physique Nucl\'eaire et de Physique des Particules in France; the Science \& Technology Facilities Council in the United Kingdom; and the Department of Energy, the National Science Foundation, and the LSST Corporation in the United States.  DESC uses resources of the IN2P3 Computing Center (CC-IN2P3--Lyon/Villeurbanne - France) funded by the Centre National de la Recherche Scientifique; the National Energy Research Scientific Computing Center, a DOE Office of Science User Facility supported by the Office of Science of the U.S.\ Department of Energy under Contract No.\ DE-AC02-05CH11231; STFC DiRAC HPC Facilities, funded by UK BIS National E-infrastructure capital grants; and the UK particle physics grid, supported by the GridPP Collaboration.  This work was performed in part under DOE Contract DE-AC02-76SF00515.

%%%%%%%%%%%%%%%%%%%%%%%%%%%%%%%%%%%%%%%%%%%%%%%%%%



\bibliographystyle{mnras}
\bibliography{Refs/references.bib}

% \input{Local_credible_regions.bbl}
%%%%%%%%%%%%%%%%%%%%%%%%%%%%%%%%%%%%%%%%%%%%%%%%%%

\appendix
\section{Regularization Marginalization} \label{sec:MM}

A prior $f(\kappa)$ is $k$-homogeneous if $\exists \; k \in \mathbb{R}_{+}$ such that
\begin{equation} \label{eq:homogeneity}
f(\eta \kappa) = \eta^kf(\kappa), \: \forall \kappa \in \mathbb{R}^n, \: \forall \eta > 0.
\end{equation}
As all norms, composite norms and compositions of norms and linear operators \citep{[16]} have homogeneity of 1, $k$ in our setting is set to 1. If we wish to infer $\kappa$ without \textit{a priori} knowledge of $\mu$ (the regularization parameter) then we calculate the normalization factor of $p(\kappa | \mu)$,
\begin{equation}
C(\mu) = \int_{\mathbb{C}^N} \exp \lbrace -\mu f(\kappa) \rbrace d\kappa.
\end{equation}
For the vast majority of cases of interest, calculating $C(\mu)$ is not feasible, due to the large dimensionality of the integral. However, it was recently shown \citep{[16]} that if the prior term $f(\kappa)$ is $k$-homogeneous then
\begin{equation} \label{eq:proposition}
C(\mu) = D \mu^{-N/k}, \quad \text{where,} \quad D \equiv C(1).
\end{equation}
A gamma-type hyper-prior is then selected (a typical choice for scale parameters) on $\mu$ such that
\begin{equation} \label{eq:hyper-prior}
p(\mu) = \frac{\beta^{\alpha}}{\Gamma(\alpha)}\mu^{\alpha - 1} e^{-\beta \mu} \mathbb{I}_{\mathbb{R}_+}(\mu),
\end{equation}
where the hyper-parameters $(\alpha, \beta)$ are very weakly dependent and can be set to 1 \citep[as in][]{[16]} and $\mathbb{I}_{C_{\alpha}}$ is an indicator function defined by
\begin{equation} \label{eq:indicator}
  \mathbb{I}_{C_{\alpha}}=\begin{cases}
               1 \quad \text{if,}\quad  \kappa \in C_{\alpha} \\
               0 \quad \text{if,}\quad \kappa \not\in C_{\alpha}.\\
            \end{cases}
\end{equation}
Now construct a joint Bayesian inference problem of $p(\kappa,\mu | \gamma)$ with MAP estimator $(\kappa^{\text{map}}, \mu^{\text{map}}) \in \mathbb{C}^{N} \times \mathbb{R}_+$. By definition, at this MAP estimator
\begin{equation}
\mathbf{0}_{N+1} \in \partial_{\kappa, \mu} \log p(\kappa^{\text{map}}, \mu^{\text{map}} | \gamma),
\end{equation}
where $\mathbf{0}_{i}$ is the $i$-dimensional null vector. This in turn implies both that
\begin{equation} \label{eq:initial-opt}
\mathbf{0}_N \in \partial_{\kappa} \log p(\kappa^{\text{map}}, \mu^{\text{map}} | \gamma),
\end{equation}
from which equation (\ref{eq:log-posterior}) follows naturally, and
\begin{equation} \label{eq:new-opt}
\mathbf{0} \in \partial_{\mu} \log p(\kappa^{\text{map}}, \mu^{\text{map}} | \gamma).
\end{equation}
Using equations (\ref{eq:proposition}, \ref{eq:hyper-prior}, \ref{eq:new-opt}) it can be shown \citep{[16]} that
\begin{equation}
\mu^{\text{map}} = \frac{\frac{N}{k} + \alpha - 1}{f(\kappa^{\text{map}}) + \beta}.
\end{equation}
Hereafter we drop the map superscript on $\mu^{\text{map}}$ for simplicity. In order to compute the MAP $\mu$ preliminary iterations are performed as follows:
\begin{equation}
\kappa^{(t)} = \argminT_{\kappa}  \big \lbrace f(\kappa ; \mu^{(t)}) + g(\kappa) \big \rbrace,
\end{equation}
\begin{equation}
\mu^{(t+1)} = \frac{\frac{N}{k}+\alpha-1}{f(\kappa^{(t)}) + \beta},
\end{equation}
where $\alpha$ and $\beta$ are (weakly dependent) hyper-parameters from a gamma-type hyper-prior, $N$ is the dimension of the reconstructed space, and the sufficient statistic $f(\kappa)$ is $k$-homogeneous. Typically the MAP solution of $\mu$ converges within $\sim 5-10$ iterations, after which $\mu$ is fixed and the optimization in equation (\ref{eq:log-posterior}) is computed.

% Don't change these lines
\bsp	% typesetting comment
\label{lastpage}
\end{document}